\documentclass[aps,prl,showpacs,superscriptaddress,twocolumn]{revtex4}

\usepackage{epsfig}
\usepackage{amsmath,amssymb,amsthm}
\usepackage{graphicx}
\usepackage{psfrag}
\usepackage{bm}

\def\s2{\sigma^2}

\begin{document}
\title{Identifying wave packet fractional revivals by means of information entropy}

\author{E. Romera}
\affiliation{Instituto Carlos I de F{\'\i}sica Te\'orica y
Computacional, Universidad de Granada, Fuentenueva s/n, 18071 Granada,
Spain}

\author{F. de los Santos}
\affiliation{Instituto Carlos I de F{\'\i}sica Te\'orica y
Computacional, Universidad de Granada, Fuentenueva s/n, 18071 Granada,
Spain}

\date{\today}

\begin{abstract}
Wave packet fractional revivals is a relevant feature in the long time scale
evolution of a wide range of physical systems, including atoms, molecules and
nonlinear systems. We show that the sum of information entropies in both position and momentum 
conjugate spaces is an indicator of fractional revivals by analyzing three different model 
systems: $(i)$ the infinite square well, $(ii)$ a particle bouncing vertically against a wall
in a gravitational field, and $(iii)$ the vibrational dynamics of hydrogen iodide molecules.
This description in terms of information entropies complements the usual one in terms of the
autocorrelation function.

\end{abstract}
\maketitle

The phenomenon of quantum wave packet revivals has received wide
attention over the last years. It has been investigated theoretically 
in atomic and molecular quantum systems \cite{1}, 
and observed experimentally in, among others, 
Rydberg wave packets in atoms and molecules, molecular vibrational states, and
Bose-Einstein condensates \cite{2}.
Revivals occur when a wave packet solution of 
the Schr\"odinger equation evolves in time to a state that closely 
reproduces its initial waveform. Fractional revivals
appear as the temporal formation of structures that are given by a
superposition of shifted and rephased initial wave packets \cite{3,4,rob}. 
It has been shown \cite{parker,3} that the relevant time scales of wave
function evolution are contained in the coefficients of the
Taylor series of the energy spectrum, $E_n$, around the energy
$E_{n_0}$ corresponding to the peak of the initial wave packet. More precisely,
the second-, third-, and fourth-order terms in this expansion 
are associated with, respectively, the classical period of motion $T_{cl}$, 
the quantum revival time-scale $T_{rev}$, and the so-called superrevival time.
Fractional revival times can be given in
terms of the quantum revival time-scale by $t=p T_{rev}/q$,
with $p$ and $q$ mutually prime \cite{3},
and are usually analyzed using the
autocorrelation function $A(t)$, which is the overlap between the initial and the 
time-evolving wave packet \cite{primenumbers}. 
At certain fractional revivals, however, the autocorrelation function may be of limited help since the wave packet reforms itself, possibly into a scaled copy of its original shape, in a location that does not generally coincide with its initial position. An expectation value analysis of wave packet evolution has been recently proposed by some authors, but it misses to fully detect the fractional revivals \cite{rob,wal,sun}.

In this Letter we study the wave packet dynamics by means of the sum of the information entropies 
of the probability density of the wave packet, in both position and momentum spaces. 
We shall show that it provides a natural framework for fractional revival
phenomena. The position-space information entropy measures the uncertainty in
the localization of the particle in space, so the lower
is this entropy the more concentrated is the wave function, the
smaller is the uncertainty, and the higher is the accuracy in
predicting the localization of the particle. Momentum-space
entropy measures the uncertainty in predicting the momentum of the
particle. 
Thus, information entropy gives an account of the spreading 
(high entropy values) and the regenerating (low entropy values) of initially well localized 
wave packets during the time evolution. 
Moreover, if $\rho(x)=|\psi(x)|^2$ and $\gamma(p)=|\phi(p)|^2$ are 
respectively the probability densities in position and momentum 
spaces (where $\psi$ and $\phi$ are the position and momentum wave
packets), the uncertainty relation for the information entropy implies
$S_{\rho}+S_{\gamma}\geq 1+\ln\pi$,
where $S_{\rho}=-\int \rho(x) \ln\rho(x) d x$
and, analogously, $S_{\gamma}=-\int \gamma(p)
\ln\gamma(p) d p$. 
This inequality is a generalization of the standard variance-based
Heisenberg uncertainty relation \cite{bbm}.
It is satisfied as a strict equality only 
for Gaussian wave packets and bounds from below
the sum of the entropies to $1+\ln\pi$. 
During the evolution of a Gaussian wave
packet, the entropy sum decreases at the revival times to reach
the above value, which plays a similar role to unity in the 
autocorrelation function.
Furthermore, the formation of a number of `minipackets' of
the original packet, i.e. fractional revivals of the wave function, will correspond
to the relative minima of the total entropy. 
Notice that it is the sum of the entropies that is employed
as an indicator of the fractional revivals, 
and not either of them separately,
because only the sum embraces
both the configurational and the motion aspects of the wave packet dynamics.
In this context, we point out that the sum of entropies for the phase and 
photon number has been used to study the formation of macroscopic quantum 
superposition states from an initially coherent state via interaction with  
a Kerr medium \cite{orlowski}. 
Additionally, we note that a finite difference eigenvalue method has been recently 
derived from which the various orders of revivals can be directly calculated 
rather than searching for them \cite{Laser}.

In what follows we investigate the time evolution of
wave packets in three one-dimensional, model systems that
exhibit fractional revival behavior, namely, the familiar infinite square-well, the
quantum 'bouncer', that is, a quantum particle bouncing on a hard surface 
under the influence of gravity, and a superposition of molecular wave packets
in a Morse potential describing the vibrations of hydrogen iodide molecules.
The infinite square-well is especially well suited for the understanding of fractional revival
phenomena because of its amenability to analytical treatment, and has been analyzed 
by several authors (see, for instance, \cite{4,rob,rob2}).
On the other hand, the quantum bouncer is the quantum version of a familiar
classical system and has also been studied in the context of wave packet
propagation and revival phenomena \cite{chen,rob,don}. Recently, gravitational quantum
bouncers have been realized using neutrons \cite{nes} and atomic clouds \cite{bon}, 
hence endowing this system with great physical significance. 
Finally, molecular wave packets provide a realistic scenario for the assessment 
of the entropy approach. 
Revivals and fractional revivals have been observed experimentally
in wave packets involving vibrational levels and can be
probed by random-phase fluorescence interferometry \cite{2,17}

Consider an infinite potential-well defined as
$V(x)=0$ for $0<x<L$ and $V(x)=+\infty$ otherwise.
The time-dependent wave function for a localized quantum wave packet is expanded as
a one-dimensional superposition of energy eigenstates as 
\begin{equation}
\psi(x,t)=\sum_n a_n u_n(x) e^{-i E_n t/\hbar},
\label{evolutionx}
\end{equation}
where $u_n(x)$ represent the normalized eigenstates and $E_n$ the corresponding eigenvalues,
$u_n(x)=\sqrt{2/L} \sin \left(n\pi x/L\right), E_n=n^2 \hbar^2\pi^2/2 m L^2$.
Following the customary procedure,
the classical period and the revival time can be computed as
$T_{cl}=2 m L^2/\hbar \pi n$ and $T_{rev}=4mL^2/\hbar \pi$, respectively. 
As an example, 
it is easy to see by direct substitution in (\ref{evolutionx})
that $\psi(L-x,T_{rev}/2)=-\psi(x,0)$, so at time $t=T_{rev}/2$ a copy of the initial 
state reforms itself, reflected around the center of the well \cite{4}.

We shall consider an initial Gaussian wave packet with a width 
$\sigma$, centered at a position $x_0$ and with a momentum $p_0$,
$\psi(x,0)=\exp[-(x-x_0)^2/2\alpha^2\hbar^2+ip_0(x-x_0)/\hbar]/ 
\sqrt{\alpha\hbar\sqrt{\pi}}
$.
Assuming that the integration region can be extended to the whole real axis,
the expansion coefficients can be approximated with high accuracy by an analytic 
expression \cite{rob}.
\begin{figure}
\centerline{\psfig{file=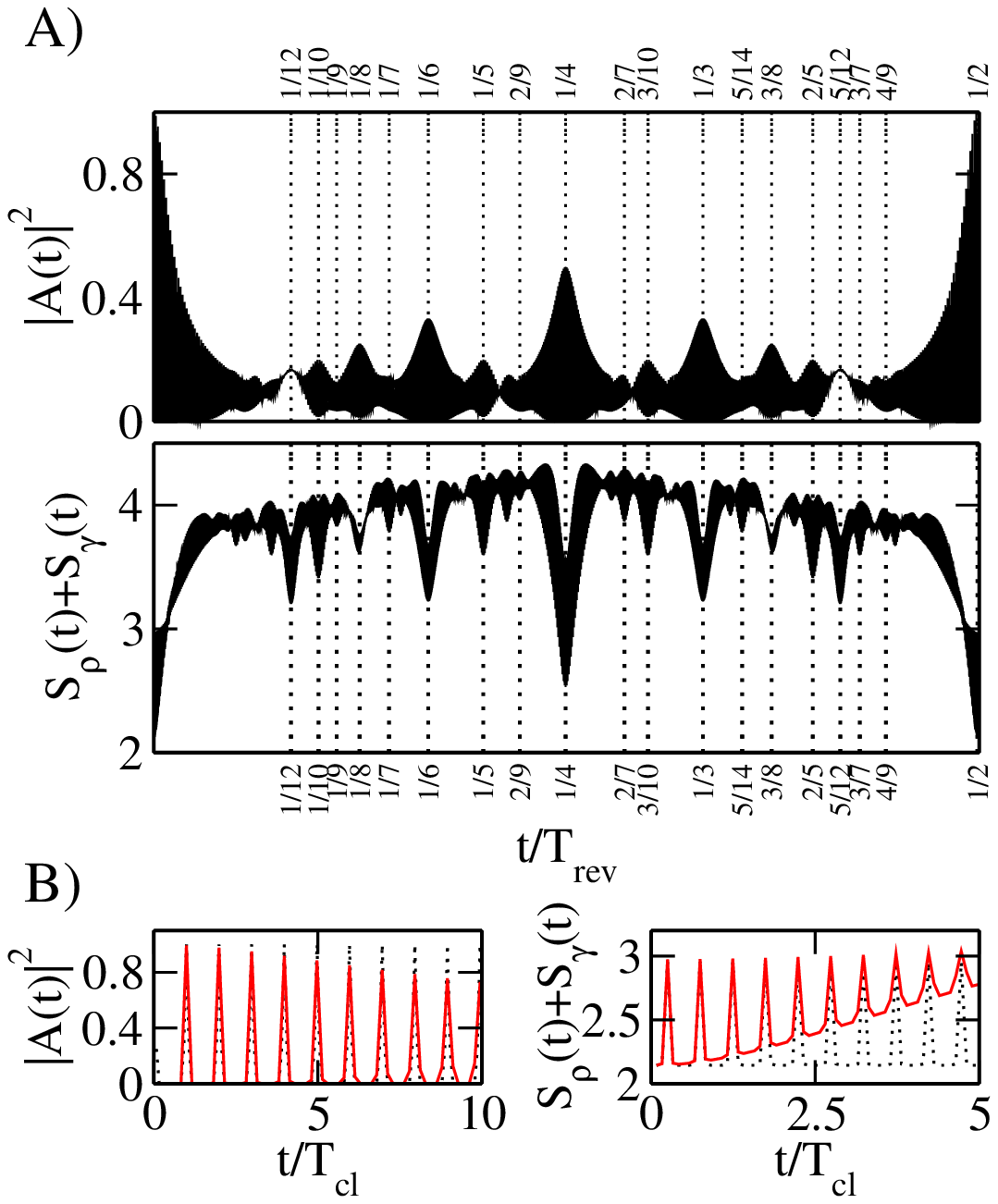,width=7.8cm,angle=0}}
\caption{
Time dependence of $|A(t)|^2$ and $S_\rho(t)+S_{\gamma}(t)$ for an 
initial Gaussian wave packet with $x_0=L/2$, $p_0=400\pi$, and
$\sigma=1/10$ in an infinite square-well.
A) Long-time dependence. The main fractional revivals are indicated by 
vertical dotted-lines.
B) First classical periods of motion. Quantum wave packets 
are represented by solid lines and their classical component by dashed-lines.}
\label{suma400}
\end{figure}
To calculate the corresponding time dependent, momentum wave function
we use the Fourier transform of the equation (\ref{evolutionx}), 
and the momentum-space normalized eigenstates 
\begin{equation}
\phi_n(p)=\sqrt{\frac{\hbar}{\pi L}}
\frac{p_n}{p^2-p^2_n}\bigg[(-1)^n e^{ipL/\hbar}-1\bigg].
\end{equation}
Without loss of generality, we shall henceforth take $2m=\hbar=L=1$, 
and $\sigma=1/10$ for the initial wave packet.

In Fig. \ref{suma400} the sum of entropies, $S_\rho(t)+S_\gamma(t)$, and the
autocorrelation function, $|A(t)|$, are shown for an initial
wave packet with $x_0=0.5$ and $p_0=400\pi$. At early times, 
the Gaussian wave packet evolves quasi-classically, but in a few periods
the quantum and classical wave packet trajectories start moving apart 
(Fig. \ref{suma400}B), the classical component of the wave 
function being defined as $\psi_{cl}(x,t)=\sum_n a_n u_n(x) e^{-i 2\pi n t/T_{cl}}$ \cite{rob}. 
For longer time scales, a large amplitude modulation is superimposed on the quasiperiodic
oscillations (Fig. \ref{suma400}A). In this long-time regime, 
the wave packet initially spreads and delocalizes while undergoing a sequence 
of fractional revivals with the creation of sub-packets, 
each of them similar to the initial one so that the sum of the entropies reaches a relative minimum
and the modulus of the autocorrelation function a relative maximum.
The most important fractional revivals are denoted by vertical dashed-lines. 
It can be observed that the identification of some fractional revivals from 
the autocorrelation function is not so clear-cut as compared with the entropy approach
(see, for example, the cases $t=p T_{rev}/q$  with $q=7$ or $9$).

\begin{figure}
\centerline{\psfig{file=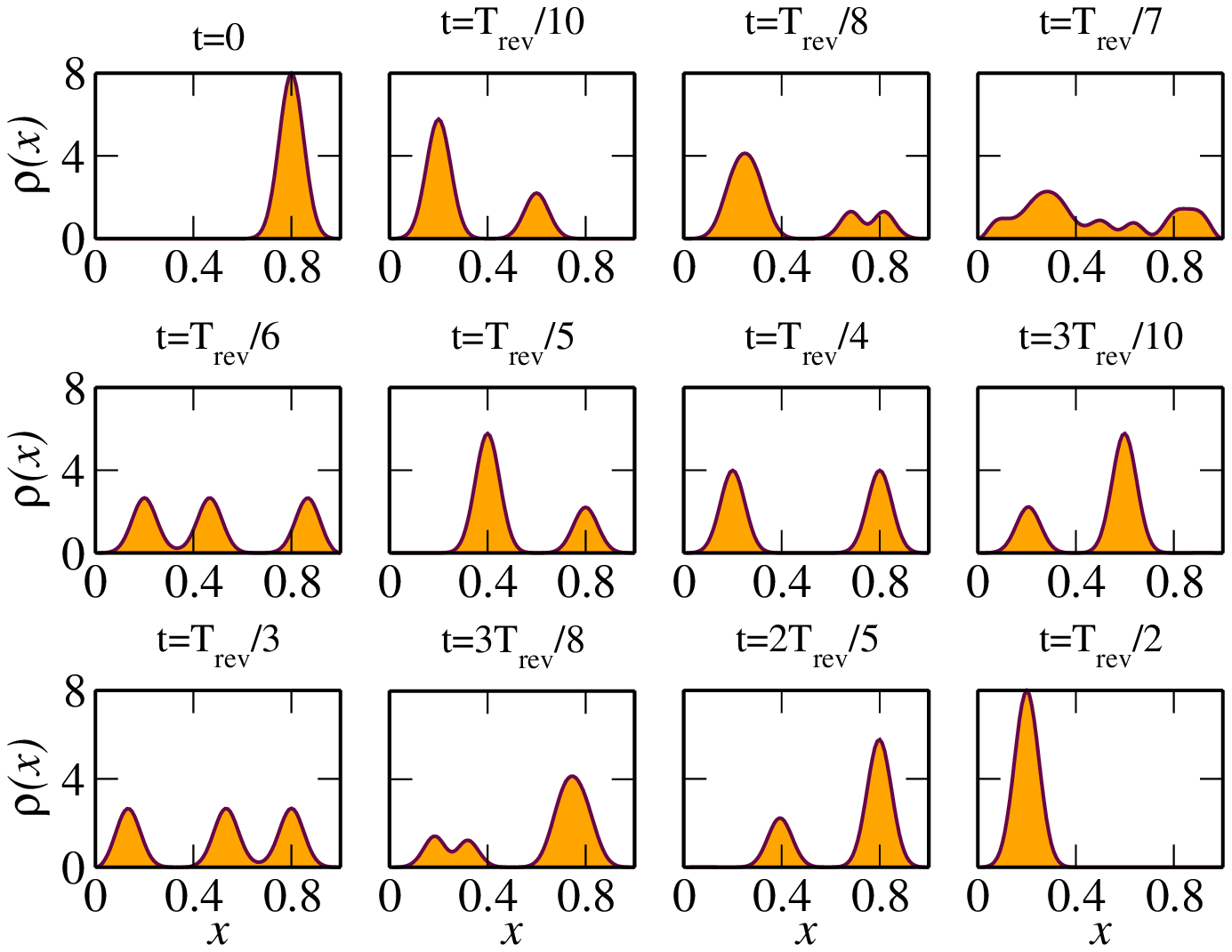,height=7.8cm,angle=-90}}
\caption{Snapshots of the probability density in position-space for 
an initial Gaussian wave packet with $x_0=0.8 L$, $p_0=0$, and
$\sigma=1/10$ in an infinite square-well and at different fractional
revival times.}
\label{snapshots}
\end{figure}

\begin{figure}
\centerline{\psfig{file=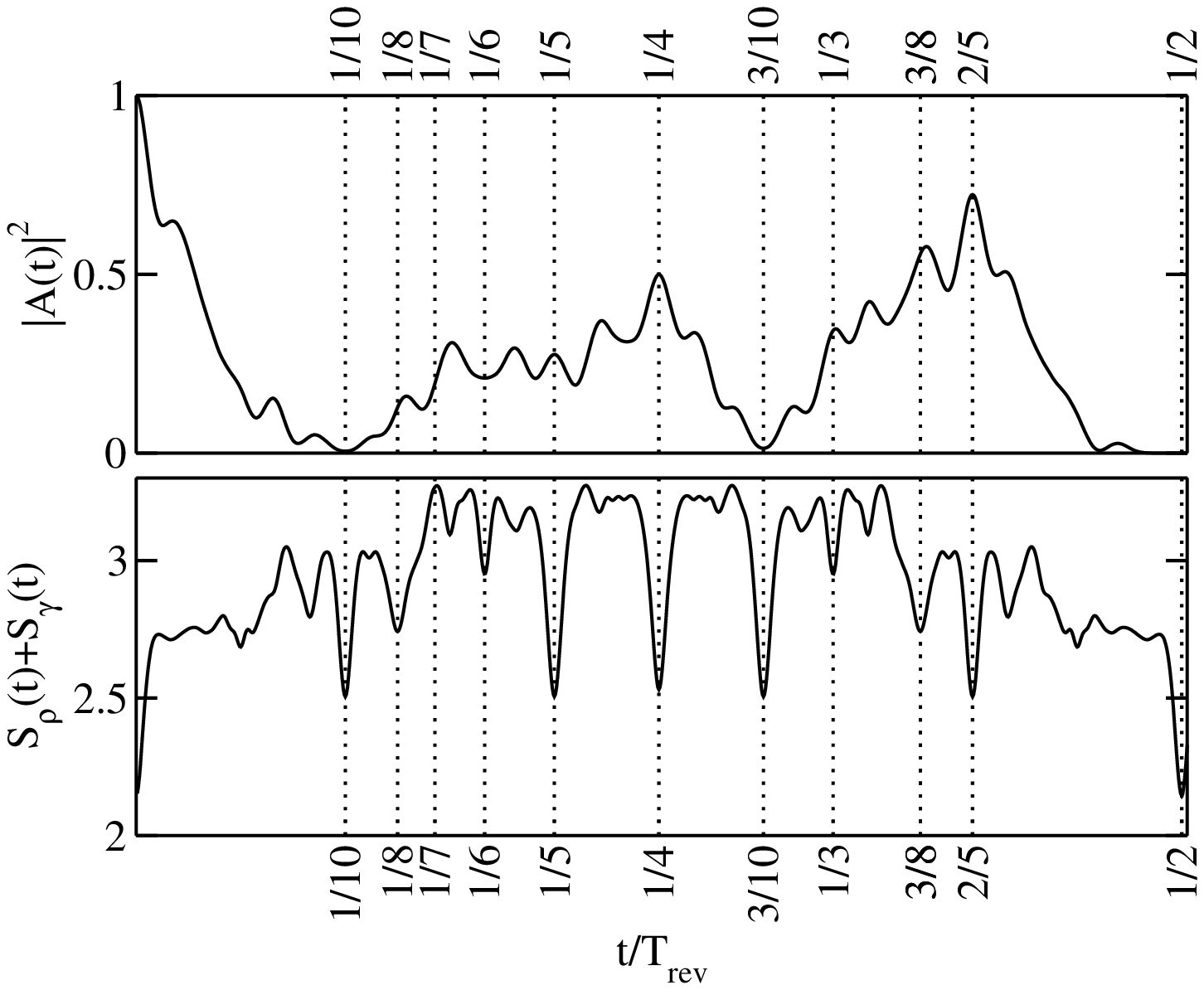,height=7.8cm,angle=-90}}
\caption{Time dependence of (top panel) $|A(t)|^2$ and
(bottom panel) $S_\rho(t)+S_{\gamma}(t)$ for an initial Gaussian 
wave packet in an infinite square well. Parameters as in Fig. \ref{snapshots}.}
\label{suma0}
\end{figure}

We have also investigated the interesting initial condition $x_0=0.8L$ and $p_0=0$, 
for which there is no classical periodic motion.
Snapshots of the numerical simulation of the position-space
probability density are given in Fig. \ref{snapshots} at several times. 
It is apparent from the bottom panel of Fig. \ref{suma0} that the sum of entropies has a minimum 
at the main fractional revivals, denoted by vertical dashed-lines, except for the case
$t=T_{rev}/7$ where the position-space probability density has a shape
compatible with a collapsed wave function (Fig. \ref{snapshots}). 
The autocorrelation function, as plotted in the top panel of Fig. \ref{suma0}, fails to show the 
fractional revivals occurring at, for example, $t/T_{rev}=1/6$, $1/8$, and $3/10$.
This discrepancy can be traced back to the fact that information entropies take into account individual minigaussian packets independently of their  relative positions, whereas the autocorrelation function 
depends on the relative position between the initial wave packet and the evolved one.

Next, we study the behavior of these same quantities in other physical situations.
Consider a particle bouncing on a hard surface under the
influence of gravity, that is, a particle in a potential
$V(z)=mgz$, if  $z>0$ and  $V(z)=+\infty$ otherwise.
Upon introducing the characteristic gravitational length
$l_g=\left(\hbar/2gm^2\right)^{1/3}$ and defining
$z^{\prime}=z/l_g$ and $E^{\prime}=E/mgl_g$, as in \cite{gea},
the eigenfunctions and eigenvalues are given by 
$E^{\prime}_n=z_n, u_n(z^{\prime})= {\cal N}_n
Ai(z^{\prime}-z_n)$ with  $n=1,2,3,\ldots$,
where $Ai(z)$ is the Airy function,
$-z_n$ denotes its zeros, and ${\cal N}_n$ 
is the $u_n(z^\prime)$ normalization factor. 
Very accurate analytic approximations to $z_n$ and ${\cal N}_n$ can be 
found in \cite{gea}.
Consider now an initial Gaussian wave packet 
localized at a height $z_0$ above the floor, with a width $\sigma$ and an initial 
momentum $p_0=0$. The corresponding coefficients in Eq. 
(\ref{evolutionx}) can be obtained analytically \cite{gea},
and the classical period and the revival time are
$T_{cl}=2\sqrt{z_0}$ and $T_{rev}=4 z_0^2/\pi$, respectively.
\begin{figure}
\centerline{\psfig{file=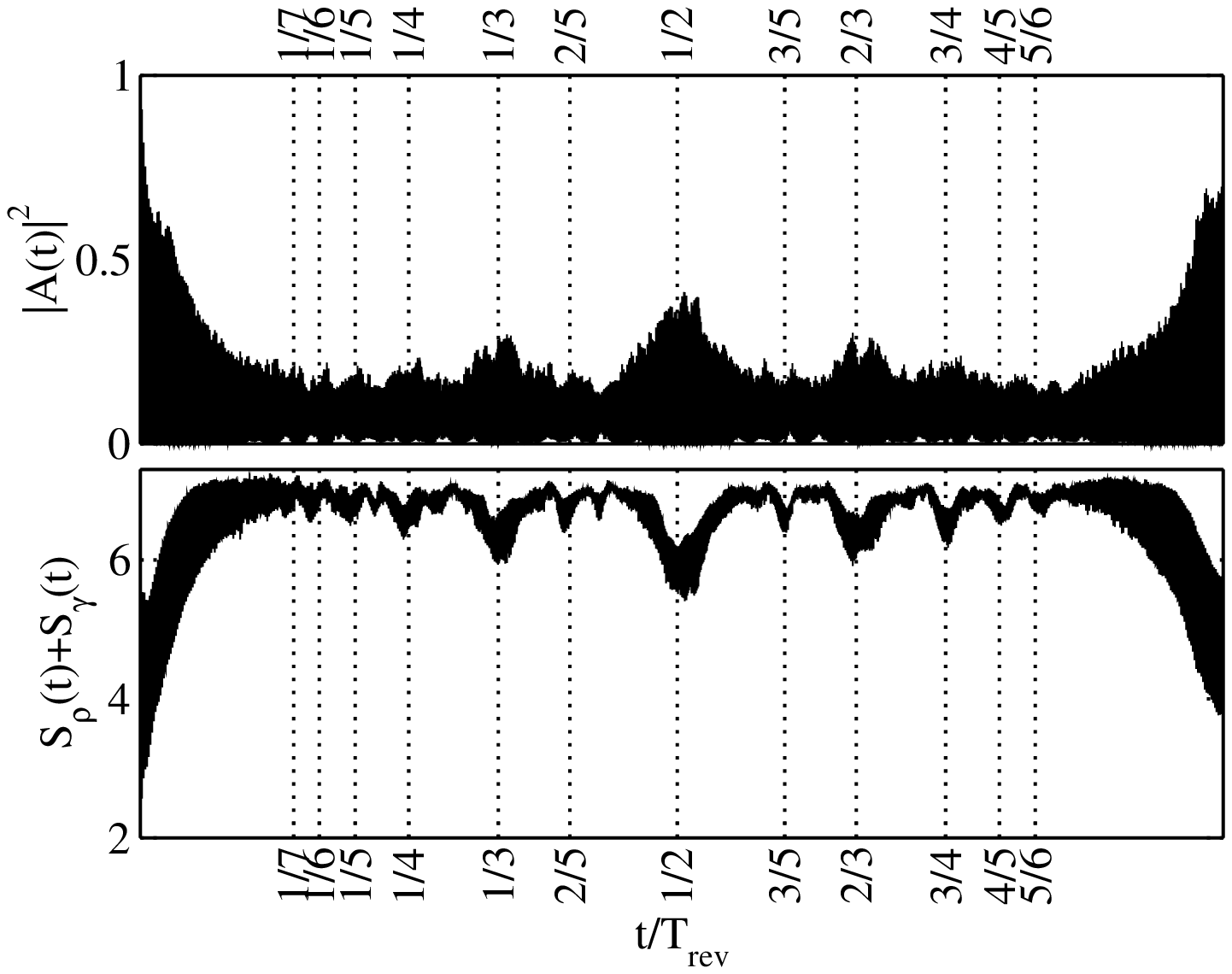,height=7.8cm,angle=-90}}
\caption{Time dependence of (top panel) $|A(t)|^2$ and (bottom panel) $S_\rho(t)+S_{\gamma}(t)$ 
for an initial Gaussian wave packet with $z_0=100$, $p_0=0$, and
$\sigma=1$ in a quantum bouncer. The main fractional revivals are indicated by vertical
dashed-lines.}
\label{sumabou}
\end{figure}

\begin{figure}
\centerline{\psfig{file=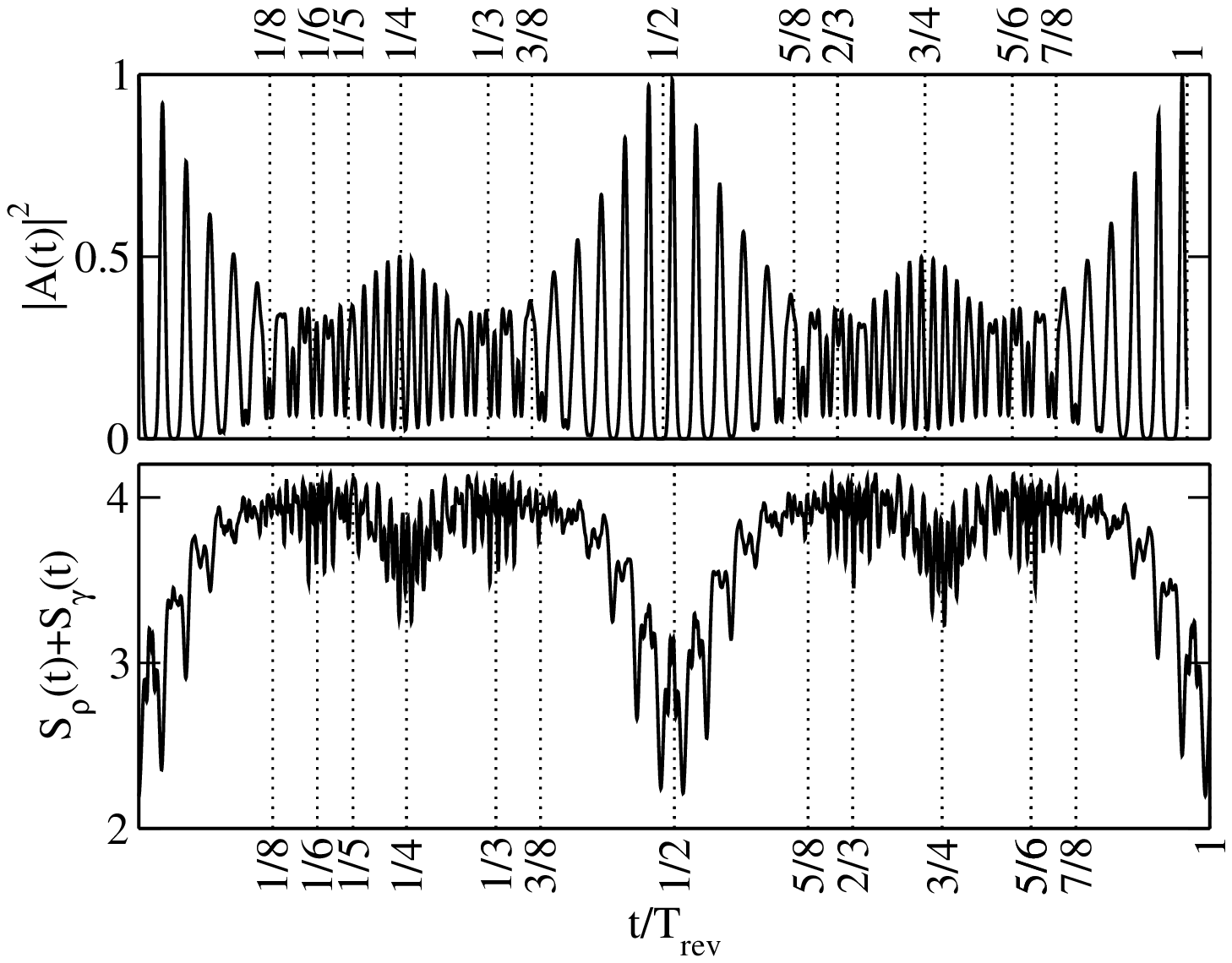,height=7.8cm,angle=-90}}
\caption{Time dependence of (top panel) $|A(t)|^2$ and (bottom panel) $S_\rho(t)+S_{\gamma}(t)$ for
an initial superposition of molecular wave packet with Gaussian weights 
in a Morse potential. The main fractional revivals are indicated by vertical
dashed-lines.}
\label{molecule}
\end{figure}

The temporal evolution of the total entropy and of 
the autocorrelation function was computed numerically for the initial
conditions $z_0=100$, $\sigma=1$, and $p_0=0$ (see Fig. \ref{sumabou}). 
For that, the corresponding wave packet in momentum-space was obtained numerically
by the fast Fourier transform method. One can see in the bottom panel of Fig. \ref{sumabou}
that the reformation of subpackets at fractional revivals is captured by the successive
relative minima of $S_\rho(t)+S_\gamma(t)$. 
A similar information is provided by the, somewhat less clear, sequence of relative maxima 
of $|A(t)|^2$ (top panel of Fig. \ref{sumabou}).

Last, we address the case of a diatomic molecule, the vibrational dynamics of which
is known to be well approximated by the Morse potential, $V(x)=D(e^{-2\beta x}-2e^{-\beta x})$.
Here, $x=r/r_0-1$, $r$ is the internuclear distance, $r_0$ is the equilibrium bond distance, 
$D$ is the dissociation energy and $\beta$ is a range parameter.
Defining $\lambda= \sqrt{2\mu D}r_0/\beta\hbar$ and $s=2\lambda\sqrt{-E/D}$, the associated eigenvalues and eigenfunctions for bounded states can be written as \cite{ghosh}
$\psi_n^\lambda(\xi)=N e^{-\xi/2}\xi^{s/2} L_n^s(\xi), \quad E_n=-D(\lambda-n-1/2)^2/\lambda^2$,
with $\xi=2\lambda e^{-\beta x}$, $0<\xi<\infty$, and $n=0,1,...,[\lambda-1/2]$, $[x]$ being the integer
part of $x$. $L_n^s(\xi)$ are the Laguerre polynomials and $N$ is a normalization factor 
(see \cite{ghosh}).

We shall consider the HI molecule, for which $\beta=2.07932$, $D=0.1125$ a.u., $r_0=3.04159$ a.u., with the reduced mass $\mu=1819.99$ a.u. The number of bound states is 
$[\lambda -1/2]+1=30$,  and the classical period and the revival time are given by 
$T_{cl}=T_{rev}/(2\lambda -1)$ and $T_{rev}=2\pi \lambda^2/D$, respectively.
Notice that in this case the autocorrelation function turns out to be symmetric about
$T_{rev}/2$ \cite{vetchinkin}.
As an initial wave packet, we shall take a superposition of Morse eigenstates
assuming a gaussian population of vibrational levels 
$
|c_n|^2=\exp\left((n-n_0)^2/\sigma\right)/(\pi \sigma)^{1/2},
$
with $n_0=7$, $\sigma=3$. As in the quantum bouncer case, the fast Fourier
method is used to compute the entropy in the momentum space.
These system parameters lead to the entropy and autocorrelation function
depicted in Fig. \ref{molecule}, where the occurrance of fractional revivals 
at $t=1/6, 1/3, 2/3$, and $5/6$ is transparent 
at first sight in the bottom panel, 
in contrast with the information provided by the autocorrelation function (top panel).

In conclusion, we have found that the manifestation of fractional revivals in
the long time-scale evolution of quantum wave packets 
reflects in the sum of information entropies in conjugate spaces, which clearly shows 
relative minima at the fractional revival times, 
thus providing a useful tool to reckon with for visualizing 
fractional revivals, complementary to the conventional autocorrelation function. 
These minima appear independently of the relative position of the subpackets 
that configure the wave packet at the fractional revival time, and are rigorously bounded 
from below. It would be very interesting to carry out  a more systematic
study extending this approach to systems with two or more quantum numbers.

This work was supported by the
Spanish projects FIS2005-00791/00973 and
FQM-165/0207.

\end{document}